\documentclass[11pt,english,onecolumn]{IEEEtran}
\usepackage[T1]{fontenc}
\usepackage[latin9]{inputenc}
\usepackage{color}
\usepackage{babel}
\usepackage{amsmath}
\usepackage{amsthm}
\usepackage{amssymb}
\usepackage{undertilde}
\usepackage{graphicx}
\usepackage[unicode=true,pdfusetitle,
 bookmarks=true,bookmarksnumbered=true,bookmarksopen=true,bookmarksopenlevel=3,
 breaklinks=false,pdfborder={0 0 1},backref=false,colorlinks=true]
 {hyperref}
\hypersetup{
 pdfborderstyle=,pdfborderstyle={},pdfborderstyle={},pdfborderstyle={},pdfborderstyle={},pdfborderstyle={},pdfborderstyle={},pdfborderstyle={},pdfborderstyle={},pdfborderstyle={},pdfborderstyle={},pdfborderstyle={},pdfborderstyle={},pdfborderstyle={},pdfborderstyle={},pdfborderstyle={},pdfborderstyle={},pdfborderstyle={},pdfborderstyle={},pdfborderstyle={},pdfpagelayout=OneColumn,pdfnewwindow=true,pdfstartview=XYZ,plainpages=false,linkcolor=blue,urlcolor=blue,citecolor=red,anchorcolor=blue,linkcolor=blue,urlcolor=blue,citecolor=red,anchorcolor=blue}

\makeatletter

\providecommand{\tabularnewline}{\\}

\theoremstyle{plain}
\newtheorem{thm}{\protect\theoremname}
\theoremstyle{plain}
\newtheorem{lem}{\protect\lemmaname}
\theoremstyle{plain}
\newtheorem{cor}{\protect\corollaryname}
\theoremstyle{plain}
\newtheorem{fact}{\protect\factname}
\theoremstyle{remark}
\newtheorem{rem}{\protect\remarkname}

\usepackage{babel}
\usepackage{accents} 
\usepackage{enumerate}
\usepackage{verbatim}
\usepackage{mathrsfs}
\usepackage{cite}

\usepackage[T3,T1]{fontenc}
\DeclareSymbolFont{tipa}{T3}{cmr}{m}{n}
\DeclareMathAccent{\invbreve}{\mathalpha}{tipa}{16}

\def\undertilde#1{\mathord{\vtop{\ialign{##\crcr
$\hfil\displaystyle{#1}\hfil$\crcr\noalign{\kern1.5pt\nointerlineskip}
$\hfil\tilde{}\hfil$\crcr\noalign{\kern1.5pt}}}}}

\newcommand{\subsetsim}{\mathrel{%
  \ooalign{\raise0.2ex\hbox{$\subset$}\cr\hidewidth\raise-0.8ex\hbox{\scalebox{0.9}{$\sim$}}\hidewidth\cr}}}
\newcommand{\supsetsim}{\mathrel{%
  \ooalign{\raise0.2ex\hbox{$\supset$}\cr\hidewidth\raise-0.8ex\hbox{\scalebox{0.9}{$\sim$}}\hidewidth\cr}}}

\newcommand{\subsetapprox}{\mathrel{%
  \ooalign{\raise0.4ex\hbox{$\subset$}\cr\hidewidth\raise-0.8ex\hbox{\scalebox{0.9}{$\approx$}}\hidewidth\cr}}}
\allowdisplaybreaks[1]
\def\1{\mathbf{1}}

\providecommand{\corollaryname}{Corollary}
\providecommand{\lemmaname}{Lemma}

\providecommand{\remarkname}{Remark}
\providecommand{\theoremname}{Theorem}

\makeatother

\providecommand{\corollaryname}{Corollary}
\providecommand{\factname}{Fact}
\providecommand{\lemmaname}{Lemma}
\providecommand{\remarkname}{Remark}
\providecommand{\theoremname}{Theorem}

\begin{document}
\title{The Convexity and Concavity of Envelopes of the Minimum-Relative-Entropy
Region for the DSBS}
\author{Lei Yu\thanks{L. Yu is with the School of Statistics and Data Science, LPMC \& KLMDASR,
Nankai University, Tianjin 300071, China (e-mail: leiyu@nankai.edu.cn).}}
\maketitle
\begin{abstract}
In this paper, we prove that for the doubly symmetric binary distribution,
the lower increasing envelope and the upper envelope of the minimum-relative-entropy
region are respectively convex and concave. We also prove that another
function induced the minimum-relative-entropy region is concave. These
two envelopes and this function were previously used to characterize
the optimal exponents in strong small-set expansion problems and strong
Brascamp--Lieb inequalities.  The results in this paper, combined
with the strong small-set expansion theorem derived by Yu, Anantharam,
and Chen (2021), and the strong Brascamp--Lieb inequality derived
by Yu (2021), confirm positively Ordentlich--Polyanskiy--Shayevitz's
conjecture on the strong small-set expansion (2019) and Polyanskiy's
conjecture on the strong Brascamp--Lieb inequality (2016). The proofs
in this paper are based on the equivalence between the convexity of
a function and the convexity of the set of minimizers of its Lagrangian
dual.
\end{abstract}


\section{Introduction}

Consider a doubly symmetric binary distribution $P_{XY}$ with correlation
$\rho\in(0,1)$, i.e., 
\begin{equation}
P_{XY}=\begin{array}{ccc}
X\backslash Y & 0 & 1\\
0 & \frac{1+\rho}{4} & \frac{1-\rho}{4}\\
1 & \frac{1-\rho}{4} & \frac{1+\rho}{4}
\end{array}.\label{eq:DSBS}
\end{equation}
Denote $k=\left(\frac{1+\rho}{1-\rho}\right)^{2}$. Define\footnote{Throughout this paper, the bases of all logarithms are set to $2$. }
\begin{align*}
D_{2}\left(a\right) & :=D\left(\left(a,1-a\right)\|P_{X}\right)=1-H_{2}\left(a\right),\\
D_{2}^{(a,b)}\left(p\right) & :=D\left(\begin{bmatrix}1+p-a-b & b-p\\
a-p & p
\end{bmatrix}\|P_{XY}\right),\\
\mathbb{D}_{2}\left(a,b\right) & :=\min_{0,a+b-1\le p\le a,b}D_{2}^{(a,b)}\left(p\right)\\
 & =D_{2}^{(a,b)}\left(p_{a,b}^{*}\right),
\end{align*}
where $D\left(Q\|P\right)$ denotes the relative entropy from $Q$
to $P$, $H_{2}:t\in[0,1]\mapsto-t\log t-(1-t)\log(1-t)$ is the binary
entropy function, and 
\[
p_{a,b}^{*}=\frac{\left(k-1\right)\left(a+b\right)+1-\sqrt{\left(\left(k-1\right)\left(a+b\right)+1\right)^{2}-4k\left(k-1\right)ab}}{2\left(k-1\right)}.
\]
Define the \emph{minimum-relative-entropy region} of $P_{XY}$ as
\[
\mathcal{D}\left(P_{XY}\right):=\bigcup_{a,b\in[0,1]}\left\{ \left(D_{2}\left(a\right),D_{2}\left(b\right),\mathbb{D}_{2}\left(a,b\right)\right)\right\} .
\]
Denote $H_{2}^{-1}$ as the inverse of the restriction of the binary
entropy function $H_{2}$ to the set $\left[0,\frac{1}{2}\right]$.
Denote $D_{2}^{-1}\left(s\right):=H_{2}^{-1}\left(1-s\right)$ which
is the inverse of $D_{2}$. Then, the lower and upper envelopes of
$\mathcal{D}\left(P_{XY}\right)$ respectively are 
\begin{align*}
\varphi\left(s,t\right) & =\mathbb{D}_{2}\left(D_{2}^{-1}\left(s\right),D_{2}^{-1}\left(t\right)\right)\\
\psi\left(s,t\right) & =\mathbb{D}_{2}\left(D_{2}^{-1}\left(s\right),1-D_{2}^{-1}\left(t\right)\right)\\
\varphi_{q}\left(s\right) & =\min_{0\le t\le1}\varphi\left(s,t\right)-\frac{t}{q}\\
\psi_{q}\left(s\right) & =\max_{0\le t\le1}\psi\left(s,t\right)-\frac{t}{q}.
\end{align*}
Define the lower and upper increasing envelopes of $\mathcal{D}\left(P_{XY}\right)$
respectively as 
\begin{align}
\widetilde{\varphi}\left(\alpha,\beta\right) & =\min_{s\ge\alpha,t\ge\beta}\varphi\left(s,t\right)\label{eq:lce}\\
\widetilde{\psi}\left(\alpha,\beta\right) & =\max_{s\leq\alpha,t\leq\beta}\psi\left(s,t\right)\label{eq:uce}\\
\widetilde{\varphi}_{q}\left(\alpha\right) & =\min_{s\ge\alpha}\varphi_{q}\left(s\right),\;q\ge1\\
\widetilde{\psi}_{q}\left(\alpha\right) & =\begin{cases}
\max_{s\leq\alpha}\varphi_{q}\left(s\right) & q<0\\
\max_{s\le\alpha}\psi_{q}\left(s\right) & 0<q<1
\end{cases}.
\end{align}
Denote $\underline{\Theta},\underline{\Theta}_{q}$ respectively as
the lower convex envelopes of $\widetilde{\varphi},\widetilde{\varphi}_{q}$,
and $\overline{\Theta},\overline{\Theta}_{q}$ respectively as the
upper concave envelopes of $\widetilde{\psi},\widetilde{\psi}_{q}$.
In fact, $\underline{\Theta}$ and $\overline{\Theta}$ are the optimal
exponents in the (forward and reverse) small-set expansion problems,
and they, together with $\underline{\Theta}_{q'},\overline{\Theta}_{q'}$,
are the optimal exponents in the (forward and reverse) strong Brascamp--Lieb
inequalities \cite{yu2021Graphs,yu2021strong}. Note that $q$ in
this paper in fact corresponds to its Hölder conjugate $q'$ in \cite{yu2021strong}. 

 We have the following properties of $\underline{\Theta}$, $\overline{\Theta}$,
and $\overline{\Theta}_{q}$, which implies that the directional gradients
of $\underline{\Theta}$ along the x-axis and y-axis are both not
greater than $1$, and those of $\overline{\Theta}$ and that of $\overline{\Theta}_{q}$
are all not smaller than $1$. 
\begin{lem}
\cite{yu2021strong}\label{lem:Theta} For all $\alpha,\beta\in\left[0,1\right]$
and $0\le s\leq1-\alpha,0\le t\leq1-\beta$, we have 
\begin{align}
\underline{\Theta}\left(\alpha+s,\beta+t\right)-\underline{\Theta}\left(\alpha,\beta\right) & \leq s+t\label{eq:sublinear}\\
\overline{\Theta}\left(\alpha+s,\beta+t\right)-\overline{\Theta}\left(\alpha,\beta\right) & \geq s+t\label{eq:sublinear-1}\\
\overline{\Theta}_{q}\left(\alpha+s\right)-\overline{\Theta}_{q}\left(\alpha\right) & \geq s\;\textrm{ for }q<0.\label{eq:sublinear-1-1}
\end{align}
 
\end{lem}
Moreover, it has been shown that $\widetilde{\psi}=\psi$ which means
that $\psi$ is nondecreasing. 
\begin{lem}
\label{lem:monotonicity}\cite{yu2021Graphs} We have $\widetilde{\psi}=\psi$. 
\end{lem}

Similarly, we also have the following lemma, i.e., $\invbreve\varphi_{q}$
is nondecreasing. 
\begin{lem}
\label{lem:Theta-1}\cite{yu2021strong} For $q<0$, we have $\overline{\Theta}_{q}=\invbreve\varphi_{q}.$
\end{lem}
We next introduce the main results in this paper. 
\begin{thm}
\label{thm:convexity} $\widetilde{\varphi}$ is convex on $[0,1]^{2}$. 
\end{thm}
\begin{thm}
\label{thm:concavity} $\psi$ is concave on $[0,1]^{2}$. 
\end{thm}
\begin{thm}
\label{thm:concavity2}  For $q<0$, $\varphi_{q}$ is concave on
$[0,1]$. 
\end{thm}
The proofs of Theorems \ref{thm:convexity}-\ref{thm:concavity2}
are respectively given in Sections \ref{sec:Proof-of-Theorem}-\ref{sec:Proof-of-Theorem-1-1}.
These proofs are based on the equivalence between the convexity of
a function and the convexity of the set of minimizers of its Lagrangian
dual; see Lemma \ref{lem:convexity}. Note that a more common way
to prove convexity of a function is based on the following equivalence:
 A twice differentiable function of several variables is convex on
a convex set if and only if its Hessian matrix of second partial derivatives
is positive semidefinite on the interior of the convex set. Compared
with this common equivalence, the equivalence used in this paper sometimes
is easier to verify, especially, for a function whose second derivative
is complicated. To verify the convexity of the set (or usually the
uniqueness) of minimizers of its Lagrangian dual, it usually suffices
to check the stationary points of the Lagrangian dual, which only
involves the first derivative of the function. Furthermore, when we
verify the uniqueness of the minimizer, sometimes by changing variables,
one can convert the minimization of the Lagrangian dual to a (strictly)
convex optimization problem, and hence, the uniqueness follows directly. 

By Theorems \ref{thm:convexity} and \ref{thm:concavity}, as well
as Lemma \ref{lem:monotonicity}, we know that $\underline{\Theta}=\widetilde{\varphi}$
and $\overline{\Theta}=\psi$. This, combined with the strong small-set
expansion theorem \cite{yu2021Graphs,yu2021strong}, resolves Ordentlich--Polyanskiy--Shayevitz's
conjecture on the strong small-set expansion \cite{ordentlich2020note}.
  It is easy to check that for $q\ge1$, $\varphi_{q}\left(s\right)=\min_{0\le t\le1}\widetilde{\varphi}\left(s,t\right)-\frac{t}{q}$.
Hence, $\varphi_{q}$ is nondecreasing, which implies $\widetilde{\varphi}_{q}=\varphi_{q}$.
By Theorems \ref{thm:convexity} and \ref{thm:concavity}, we have
the following corollary. 
\begin{cor}
For $q\ge1$, $\widetilde{\varphi}_{q}=\varphi_{q}$ is convex; and
for $q\in(-\infty,0)\cup(0,1)$, $\widetilde{\psi}_{q}$ is concave. 
\end{cor}

This, combined with the strong Brascamp--Lieb inequality \cite[Corollary 7]{yu2021strong},
independently resolves Polyanskiy's conjecture on strong Brascamp--Lieb
inequality stated in \cite{kirshner2019moment}. Note that as mentioned
in \cite{kirshner2019moment}, Polyansky's original conjecture was
already solved by himself in an unpublished paper \cite{polyanskiy2019hypercontractivity2}.

\begin{figure}
\centering %
\begin{tabular}{cc}
\includegraphics[width=0.5\columnwidth]{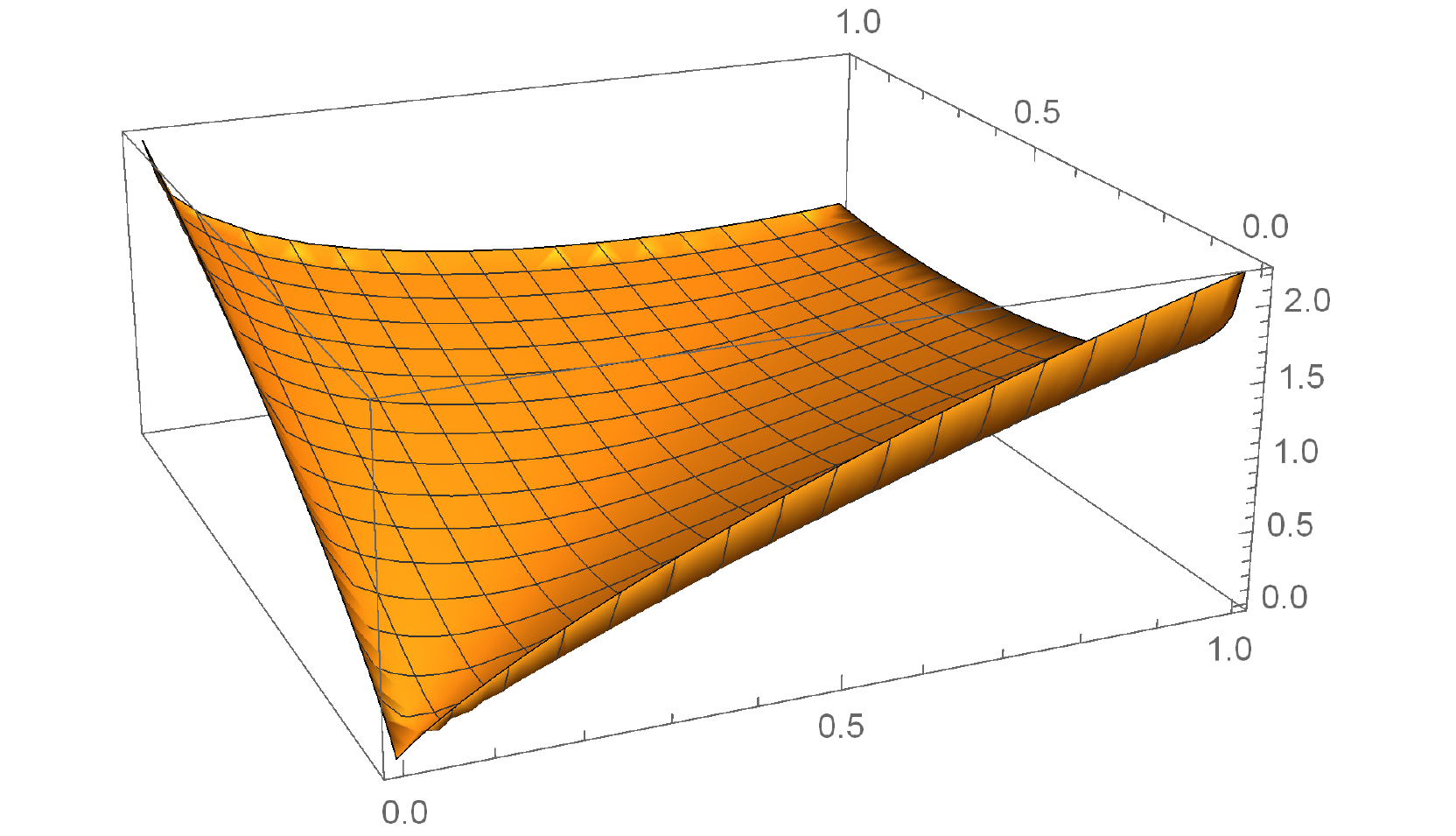} & \includegraphics[width=0.4\columnwidth,height=4cm]{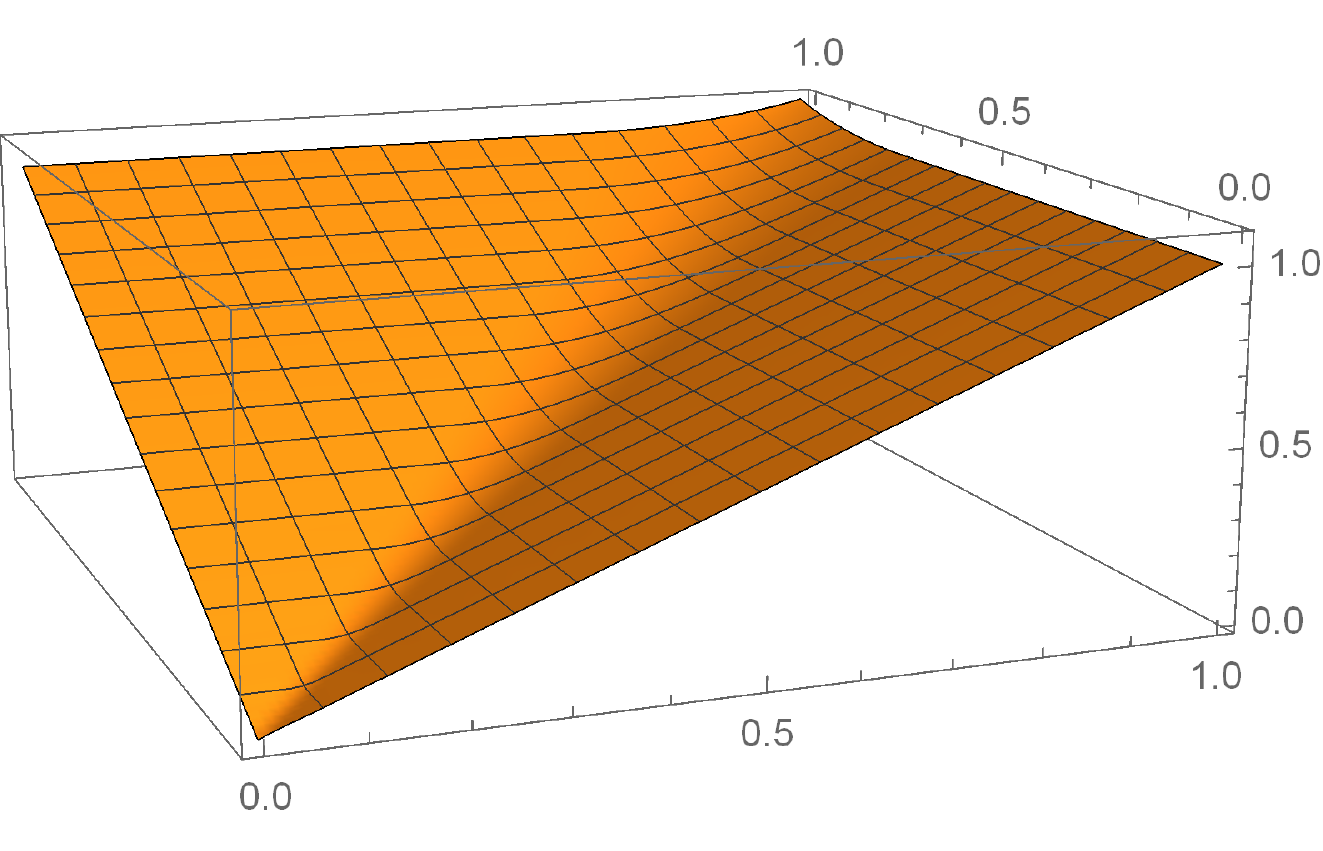}\tabularnewline
{\footnotesize{}{}$\varphi$} & {\footnotesize{}{}$\underline{\Theta}=\widetilde{\varphi}$}\tabularnewline
\end{tabular}

\begin{tabular}{cc}
\includegraphics[width=0.5\columnwidth]{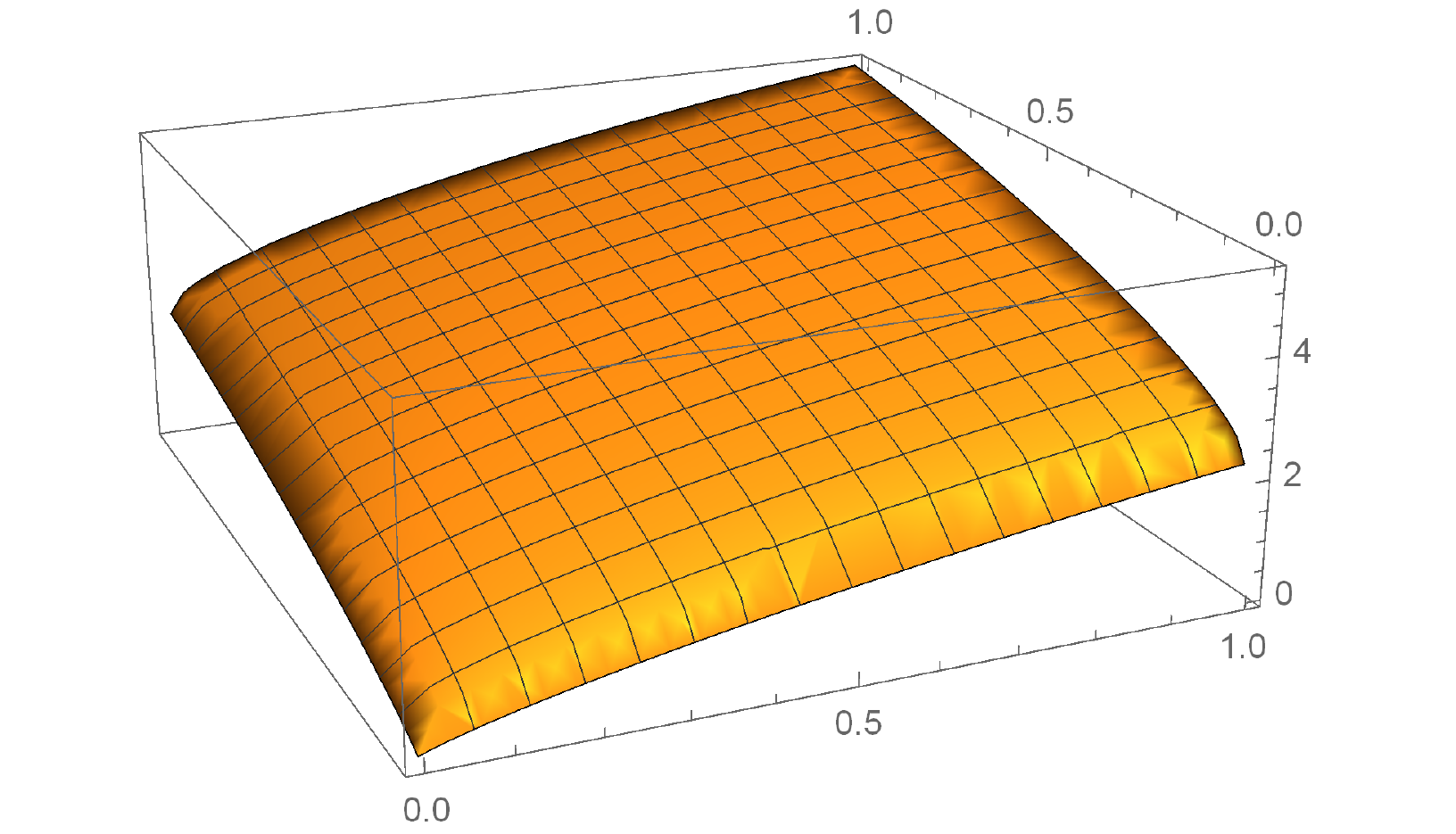} & \includegraphics[width=0.4\columnwidth]{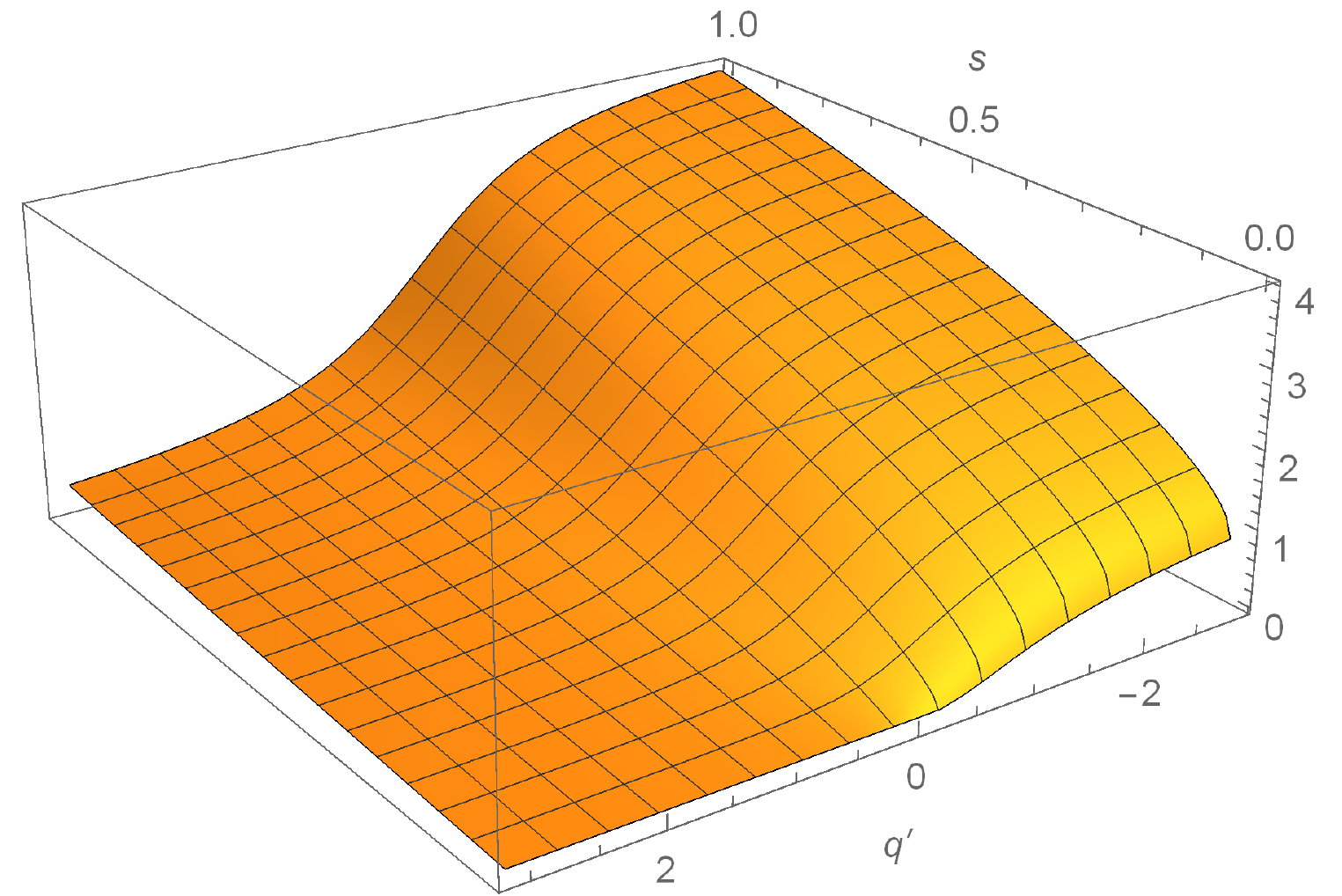}\tabularnewline
{\footnotesize{}{}$\overline{\Theta}=\widetilde{\psi}=\psi$} & {\footnotesize{}{}$\underline{\Theta}_{q}=\widetilde{\varphi}_{q}=\varphi_{q}$
for $q'\ge1$, }\tabularnewline
 & {\footnotesize{}$\overline{\Theta}_{q}=\widetilde{\psi}_{q}=\varphi_{q}$
for $0<q'<1$, }\tabularnewline
 & {\footnotesize{}$\overline{\Theta}_{q}=\widetilde{\psi}_{q}=\psi_{q}$
for $q'<0$}\tabularnewline
\end{tabular}

\caption{\label{fig:upsilon}Illustration of $\varphi,\widetilde{\varphi},\underline{\Theta},\psi,\widetilde{\psi},\overline{\Theta},$
$\varphi_{q},\widetilde{\varphi}_{q},\underline{\Theta}_{q}$, and
$\psi_{q},\widetilde{\psi}_{q},\overline{\Theta}_{q}$ for $\rho=0.9$.
In the left bottom subfigure, $\underline{\Theta}_{q}=\widetilde{\varphi}_{q}=\varphi_{q}$
(convex) for $q'\ge1$, $\overline{\Theta}_{q}=\widetilde{\psi}_{q}=\varphi_{q}$
(concave) for $0<q'<1$, and $\overline{\Theta}_{q}=\widetilde{\psi}_{q}=\psi_{q}$
(concave) for $q'<0$, where the axis $q'=\frac{q}{q-1}$ is the Hölder
conjugate of $q$.}
\end{figure}

Summerizing all the above, we have that $\widetilde{\varphi}$ and
$\varphi_{q}$ with $q\ge1$ are nondecreasing and convex;  $\psi$,
$\varphi_{q}$ with $q<0$, and $\psi_{q}$ with $0<q<1$ are nondecreasing
and concave. This further implies that 
\begin{align*}
 & \underline{\Theta}=\widetilde{\varphi}\leq\varphi,\\
 & \overline{\Theta}=\widetilde{\psi}=\psi,\\
 & \underline{\Theta}_{q}=\widetilde{\varphi}_{q}=\varphi_{q},\;q\ge1,\\
 & \overline{\Theta}_{q}=\widetilde{\psi}_{q}=\begin{cases}
\varphi_{q} & q<0\\
\psi_{q} & 0<q<1
\end{cases}.
\end{align*}
The functions $\varphi,\widetilde{\varphi},\underline{\Theta},\psi,\widetilde{\psi},\overline{\Theta},$
$\varphi_{q},\widetilde{\varphi}_{q},\underline{\Theta}_{q}$, and
$\psi_{q},\widetilde{\psi}_{q},\overline{\Theta}_{q}$ for $\rho=0.9$
are plotted in Fig. \ref{fig:upsilon}. 

\section{\label{sec:Proof-of-Theorem}Proof of Theorem \ref{thm:convexity}}

 For brevity, we denote 
\begin{align*}
f\left(\alpha,\beta\right):=\widetilde{\varphi}\left(\alpha,\beta\right) & =\inf_{s\ge\alpha,t\ge\beta}\mathbb{D}_{2}\left(D_{2}^{-1}\left(s\right),D_{2}^{-1}\left(t\right)\right)\\
 & =\inf_{a\le D_{2}^{-1}\left(\alpha\right),b\le D_{2}^{-1}\left(\beta\right)}\mathbb{D}_{2}\left(a,b\right)
\end{align*}
Observe that given $a\le1/2$, $\mathbb{D}_{2}\left(a,b\right)$ is
convex in $b$, and the minimum $D_{2}(a)$ is attained at $b=a*\frac{1-\rho}{2}\ge a$.
Here $*$ denote the binary convocation, i.e., $x*y=x(1-y)+y(1-x)$.
Hence, 
\begin{align*}
\inf_{b\le D_{2}^{-1}\left(\beta\right)}\mathbb{D}_{2}\left(a,b\right) & =\begin{cases}
D_{2}(a) & a*\frac{1-\rho}{2}\le D_{2}^{-1}\left(\beta\right)\\
\mathbb{D}_{2}\left(a,D_{2}^{-1}\left(\beta\right)\right) & a*\frac{1-\rho}{2}>D_{2}^{-1}\left(\beta\right)
\end{cases}
\end{align*}
WLOG, we assume $\alpha\ge\beta$ (i.e., $D_{2}^{-1}\left(\alpha\right)\le D_{2}^{-1}\left(\beta\right)$).
For this case, 
\begin{align}
f\left(\alpha,\beta\right) & =\inf_{a\le D_{2}^{-1}\left(\alpha\right)}\begin{cases}
D_{2}(a) & a*\frac{1-\rho}{2}\le D_{2}^{-1}\left(\beta\right)\\
\mathbb{D}_{2}\left(a,D_{2}^{-1}\left(\beta\right)\right) & a*\frac{1-\rho}{2}>D_{2}^{-1}\left(\beta\right)
\end{cases}\nonumber \\
 & =\min\left\{ \inf_{a\le D_{2}^{-1}\left(\alpha\right),a*\frac{1-\rho}{2}\le D_{2}^{-1}\left(\beta\right)}D_{2}(a),\inf_{a\le D_{2}^{-1}\left(\alpha\right),a*\frac{1-\rho}{2}>D_{2}^{-1}\left(\beta\right)}\mathbb{D}_{2}\left(a,D_{2}^{-1}\left(\beta\right)\right)\right\} \label{eq:-92}
\end{align}
We denote $*^{-1}$ as the deconvolution operation, i.e., $z*^{-1}y=\frac{z-y}{1-2y}$
is the solution to $x*y=z$ with $x$ unknown. If $D_{2}^{-1}\left(\beta\right)*^{-1}\frac{1-\rho}{2}\ge D_{2}^{-1}\left(\alpha\right)$,
\begin{align*}
f\left(\alpha,\beta\right) & =\inf_{a\le D_{2}^{-1}\left(\alpha\right),a*\frac{1-\rho}{2}\le D_{2}^{-1}\left(\beta\right)}D_{2}(a)\\
 & =\alpha
\end{align*}
If $D_{2}^{-1}\left(\beta\right)*^{-1}\frac{1-\rho}{2}<D_{2}^{-1}\left(\alpha\right)$,
then the first term in \eqref{eq:-92} satisfies that 
\[
\inf_{a\le D_{2}^{-1}\left(\alpha\right),a*\frac{1-\rho}{2}\le D_{2}^{-1}\left(\beta\right)}D_{2}(a)=D_{2}\left(D_{2}^{-1}\left(\beta\right)*^{-1}\frac{1-\rho}{2}\right)
\]
and the second term satisfies
\begin{align*}
\inf_{a\le D_{2}^{-1}\left(\alpha\right),a*\frac{1-\rho}{2}>D_{2}^{-1}\left(\beta\right)}\mathbb{D}_{2}\left(a,D_{2}^{-1}\left(\beta\right)\right) & \leq\mathbb{D}_{2}\left(D_{2}^{-1}\left(\beta\right)*^{-1}\frac{1-\rho}{2},D_{2}^{-1}\left(\beta\right)\right)\\
 & =D_{2}\left(D_{2}^{-1}\left(\beta\right)*^{-1}\frac{1-\rho}{2}\right)
\end{align*}
By the convexity of $\mathbb{D}_{2}$, the second term in \eqref{eq:-92}
also satisfies 
\begin{align*}
\inf_{a\le D_{2}^{-1}\left(\alpha\right),a*\frac{1-\rho}{2}>D_{2}^{-1}\left(\beta\right)}\mathbb{D}_{2}\left(a,D_{2}^{-1}\left(\beta\right)\right) & =\mathbb{D}_{2}\left(D_{2}^{-1}\left(\alpha\right),D_{2}^{-1}\left(\beta\right)\right)
\end{align*}
Therefore, 
\begin{align*}
f\left(\alpha,\beta\right) & =\mathbb{D}_{2}\left(D_{2}^{-1}\left(\alpha\right),D_{2}^{-1}\left(\beta\right)\right)
\end{align*}

Summarizing the above, we have 
\begin{align}
f\left(\alpha,\beta\right) & =\begin{cases}
\alpha & \left(\alpha,\beta\right)\in S_{0}\\
\beta & \left(\alpha,\beta\right)\in S_{0}^{\top}\\
\mathbb{D}_{2}\left(D_{2}^{-1}\left(\alpha\right),D_{2}^{-1}\left(\beta\right)\right) & \textrm{otherwise}
\end{cases}\label{eq:f}
\end{align}
where $S_{0}:=\left\{ \left(\alpha,\beta\right):D_{2}^{-1}\left(\beta\right)\ge D_{2}^{-1}\left(\alpha\right)*\frac{1-\rho}{2}\right\} $
and $S_{0}^{\top}:=\left\{ \left(\alpha,\beta\right):\left(\beta,\alpha\right)\in S_{0}\right\} $.
The planes $(x,y)\mapsto x$ and $(x,y)\mapsto y$ are tangent planes
of $f$ at points in $S_{0}$ and in $S_{0}^{\top}$, respectively.
Denote 

\begin{align*}
\bar{f}\left(a,b\right) & =\begin{cases}
D_{2}\left(a\right) & b\ge a*\frac{1-\rho}{2}\\
D_{2}\left(b\right) & a\ge b*\frac{1-\rho}{2}\\
\mathbb{D}_{2}\left(a,b\right) & \textrm{otherwise}
\end{cases}
\end{align*}
Note that $f$ and $\bar{f}$ are differentiable. 

Define 
\[
g\left(s,t\right):=f\left(s,t\right)-\lambda s-\mu t
\]
and 
\begin{align*}
\Gamma & :=\min_{s,t\in[0,1]}g\left(s,t\right)\\
 & =\min_{a,b\in[0,1/2]}\bar{f}\left(a,b\right)-\lambda D_{2}\left(a\right)-\mu D_{2}\left(b\right).
\end{align*}

We have the following two lemmas. 
\begin{lem}
\label{lem:minimizers}Let $S:=[0,1]^{2}$. Let $f$ be the function
given in \eqref{eq:f}, and $\breve{f}$ be its lower convex envelope.
Then, for any subgradient\footnote{If $f:S\to\mathbb{R}$ is a real-valued convex function defined on
a convex  set $S$ in the Euclidean space $\mathbb{R}^{n}$, a vector
\textbf{$\mathbf{v}$} in that space is called a \emph{subgradient}
of $f$ at a point $\mathbf{x}_{0}$ in $S^{o}$ if for any $\mathbf{x}$
in $S$ one has $f(\mathbf{x})-f(\mathbf{x}_{0})\geq\left\langle \mathbf{v},\mathbf{x}-\mathbf{x}_{0}\right\rangle $
where the $\left\langle \cdot,\cdot\right\rangle $ denotes the inner
product.  Equivalently, $\mathbf{x}\in S\mapsto f(\mathbf{x}_{0})+\left\langle \mathbf{v},\mathbf{x}-\mathbf{x}_{0}\right\rangle $
forms a supporting hyperplane of $f$ at $\mathbf{x}_{0}$. A vector
\textbf{$\mathbf{v}$} is called a \emph{supergradient} of a concave
function $g$ at a point $\mathbf{x}_{0}$ if \textbf{$-\mathbf{v}$}
is a subgradient of $-g$ at $\mathbf{x}_{0}$. } $\left(\lambda,\mu\right)$ of $\breve{f}$ at a point in $S^{o}$,
the set of minimizers of the function $g:(x,y)\mapsto f(x,y)-\lambda x-\mu y$
is  a convex subset of $S$. 
\end{lem}
\begin{lem}
\label{lem:convexity}  Let $S$ be a compact convex subset of $\mathbb{R}^{n}$,
and $f:S\to\mathbb{R}$ be a continuous function. For each point $\mathbf{y}\in S$,
let $\lambda_{\mathbf{y}}$ be a subgradient of $\breve{f}$ at $\mathbf{y}$.
For any $A\subseteq S$, let $G\left(A\right):=\left\{ \lambda_{\mathbf{y}}:\mathbf{y}\in A\right\} $.
Then, the following hold.
\begin{enumerate}
\item For any\footnote{We use $S^{o}$ to denote the interior of $S$.} $\lambda\in G\left(S^{o}\right)$,
the set of minimizers of the function $g:\mathbf{x}\in S\mapsto f(\mathbf{x})-\left\langle \lambda,\mathbf{x}\right\rangle $
is  a convex subset of $S$, if and only if   $f$ is convex on
$S$. 
\item For any $\lambda\in G\left(S^{o}\right)$, the minimizer of the function
$g:\mathbf{x}\in S\mapsto f(\mathbf{x})-\left\langle \lambda,\mathbf{x}\right\rangle $
is  unique, if and only if   $f$ is strictly convex on $S^{o}$.
\end{enumerate}
\end{lem}

By these two lemmas, we have that $f$ is convex. We next prove these
two lemmas. 

\subsection{Proof of Lemma \ref{lem:minimizers}}

Since $\breve{f}$ is increasing, any subgradient $\left(\lambda,\mu\right)$
of $\breve{f}$ must satisfy that $\lambda,\mu\ge0$. By Lemma \ref{lem:Theta},
any subgradient $\left(\lambda,\mu\right)$ of $\breve{f}$ must satisfy
that $\lambda,\mu\le1$. Hence, we only need to consider $0\le\lambda,\mu\le1$.
In the following, we denote $\lambda=1/p,\mu=1/q$ with $p,q\in[1,\infty]$. 

If $q=1,p=\infty$, then 
\[
g\left(s,t\right)=f\left(s,t\right)-t.
\]
For this case, 
\begin{align*}
\Gamma & =\min_{a,b\in[0,1/2]}\bar{f}\left(a,b\right)-D_{2}\left(b\right)\\
 & =0.
\end{align*}
This implies that the minimizers are the points in $S_{0}$. The set
$S_{0}$ is convex, since it corresponds to the set of $0\le\beta\le D_{2}\left(D_{2}^{-1}\left(\alpha\right)*\frac{1-\rho}{2}\right)$,
and $\alpha\mapsto D_{2}\left(D_{2}^{-1}\left(\alpha\right)*\frac{1-\rho}{2}\right)=1-H_{2}\left(H_{2}^{-1}\left(1-\alpha\right)*\frac{1-\rho}{2}\right)$
is concave \cite[Problem 2.5]{Gamal}. Similarly, for $p=1,q=\infty$,
the set of minimizers is also convex. 

If $1<q\le\infty,p=\infty$, then 
\[
g\left(s,t\right)=f\left(s,t\right)-\frac{t}{q}.
\]
For this case, 
\begin{align*}
\Gamma & =\min_{a,b\in[0,1/2]}\bar{f}\left(a,b\right)-\frac{1}{q}D_{2}\left(b\right)\\
 & =\min_{b\in[0,1/2]}\left(1-\frac{1}{q}\right)D_{2}\left(b\right)\\
 & =0.
\end{align*}
This implies that the minimizer of $g$ is $(0,0)$, and is unique.
 Similarly, for $1<p\le\infty,q=\infty$, the minimizer is also unique.

We now consider the case of $1\le p,q<\infty$.  Denote $Q_{XY}:=\begin{bmatrix}q_{00} & q_{01}\\
q_{10} & q_{11}
\end{bmatrix}$ as a distribution (i.e., $q_{x,y}\ge0$ for all $x,y\in\{0,1\}$
and $\sum_{x,y\in\{0,1\}}q_{x,y}=1$). Consider the following minimization
problem.
\begin{align}
\Gamma & =\min_{Q_{XY}:q_{11}+q_{10}\le1/2,q_{11}+q_{01}\le1/2}d\left(Q_{XY}\right),\label{eq:-93}
\end{align}
where 
\[
d\left(Q_{XY}\right):=D\left(Q_{XY}\|P_{XY}\right)-\frac{1}{p}D_{2}\left(q_{11}+q_{10}\right)-\frac{1}{q}D_{2}\left(q_{11}+q_{01}\right).
\]
We have the following two facts. 
\begin{fact}
If the minimizer of \eqref{eq:-93} is unique, then the minimizer
of $g$ on $[0,1]^{2}\backslash\left(S_{0}\cup S_{0}^{\top}\right)$
is also unique.  
\end{fact}
\begin{fact}
\label{fact:Lagrangian}Any strictly interior stationary points of
\eqref{eq:-93} satisfy the following Lagrangian conditions: 
\[
Q_{XY}\left(x,y\right)=\frac{\Pi\left(x,y\right)}{\sum_{x,y\in\{0,1\}}\Pi\left(x,y\right)},\forall x,y\in\{0,1\}
\]
where
\[
\Pi\left(x,y\right)=P_{XY}\left(x,y\right)\left(\frac{Q_{X}(x)}{P_{X}(x)}\right)^{1/p}\left(\frac{Q_{Y}(y)}{P_{Y}(y)}\right)^{1/q}.
\]
\end{fact}
From Fact \ref{fact:Lagrangian}, 
\begin{equation}
\left(\frac{q_{00}+q_{01}}{q_{10}+q_{11}}\right)^{1/p}=\frac{q_{00}}{q_{10}}\theta=\frac{q_{01}}{q_{11}}\frac{1}{\theta}=:z\label{eq:-90}
\end{equation}
with $\frac{q_{00}+q_{01}}{q_{10}+q_{11}}\ge1$, and 
\begin{equation}
\left(\frac{q_{00}+q_{10}}{q_{01}+q_{11}}\right)^{1/q}=\frac{q_{00}}{q_{01}}\theta=\frac{q_{10}}{q_{11}}\frac{1}{\theta}\label{eq:-91}
\end{equation}
with $\frac{q_{00}+q_{10}}{q_{01}+q_{11}}\ge1$. 

If $p=1\le q<\infty$, then solving the above equations, we have $\left(q_{00},q_{01},q_{10},q_{11}\right)=\left(\frac{1+\rho}{4},\frac{1-\rho}{4},\frac{1-\rho}{4},\frac{1+\rho}{4}\right)$.
Obviously, this stationary point is not a minimizer. Hence, the minimizers
are on the boundary $s=0$. Observe that $f\left(0,t\right)$ is convex
in $t$ for $t\ge D_{2}\left(\frac{1-\rho}{2}\right)$. Hence, the
minimizer is unique. Similarly, for $q=1\le p<\infty$, the minimizer
is also unique. 

It remains to consider the case $1<p,q<\infty$. Denote $r=\left(p-1\right)\left(q-1\right)$.
If $r>\rho^{2}$ and $1<p,q<\infty$, then by the information-theoretic
characterization of the hypercontractivity region \cite{nair2014equivalent},
$(0,0)$ is the unique minimizer of $g$. Hence, Lemma \ref{lem:minimizers}
is satisfied for this case.

We next consider the case $0<r\le\rho^{2},1<p,q<\infty$. Denote $\theta=\frac{1-\rho}{1+\rho}$,
$u:=\frac{1}{p-1},v:=\frac{1}{q-1}$. For this case, we have $u>1$
or $v>1$. By symmetry, WLOG, we assume $v>1$ here.  Solving \eqref{eq:-90}
and \eqref{eq:-91}, we have 

\begin{equation}
z^{rv}=\frac{\left(1+\theta z\right)^{v}\theta+\left(\theta+z\right)^{v}}{\left(\theta+z\right)^{v}\theta+\left(1+\theta z\right)^{v}}.\label{eq:-95}
\end{equation}

By using a proof idea similar to \cite{nair2016evaluating}, we show
the uniqueness of the root of the equation above. 
\begin{lem}
\label{lem:uniqueness}Let $|v|>1$ and $\theta\in(0,1)$. If $0<r\le\rho^{2}$,
then the equation \eqref{eq:-95} has a unique root (w.r.t. $z$)
for $z\in(1,\infty)$. 
\end{lem}

\begin{rem}
Lemma \ref{lem:uniqueness} for the case $v>1$ is sufficient to
prove Lemma \ref{lem:minimizers}. However, Lemma \ref{lem:uniqueness}
for $v<-1$ will be used to prove Theorems \ref{thm:concavity} and
\ref{thm:concavity2} in the next two sections. 
\end{rem}
\begin{IEEEproof}
Here we only focus on the case $v>1$. The case $v<-1$ follows similarly. 

Let $z=e^{h}$ and $g\left(h\right)=\log\left(\left(1+\theta e^{h}\right)^{v}\theta+\left(\theta+e^{h}\right)^{v}\right)$.
Then, the equation \eqref{eq:-95} is equivalent to 
\[
rvh=g\left(h\right)-g\left(-h\right)-hv.
\]
We define 
\[
\varphi\left(h\right)=g\left(h\right)-g\left(-h\right)-hv-rvh.
\]
Its derivative is 
\[
\varphi'\left(h\right)=g'\left(h\right)+g'\left(-h\right)-v-rv.
\]
where 
\[
g'\left(h\right)=v\left(1-\theta\left(\frac{\left(1+\theta e^{h}\right)^{v-1}+\left(\theta+e^{h}\right)^{v-1}}{\left(1+\theta e^{h}\right)^{v}\theta+\left(\theta+e^{h}\right)^{v}}\right)\right).
\]
We next show that $\varphi'\left(h\right)=0$ has a unique root on
$h>0$. 

Observe that $\varphi'\left(h\right)>0$ is equivalent to 
\begin{equation}
r\left(\eta^{v}+\eta^{-v}\right)+\eta+\eta^{-1}<\left(1-r\right)\left(\theta+\theta^{-1}\right)\label{eq:-94}
\end{equation}
where $\eta=\frac{1+\theta e^{h}}{\theta+e^{h}}$. (The equivalence
still holds if both the inequalities above change to the other direction.)

For $v>1$, $\eta\mapsto r\left(\eta^{v}+\eta^{-v}\right)+\eta+\eta^{-1}$
is convex. Hence, for $0<r\le\rho^{2}$, 
\[
r\left(\eta^{v}+\eta^{-v}\right)+\eta+\eta^{-1}=\left(1-r\right)\left(\theta+\theta^{-1}\right)
\]
has a unique root on $\eta<1$, denoted as $\eta_{0}$. Moreover,
for $\eta=1$, \eqref{eq:-94} holds. Denote $h_{0}$ such that $\eta_{0}=\frac{1+\theta e^{h_{0}}}{\theta+e^{h_{0}}}$.
Hence, $\varphi'\left(h\right)\ge0$ is equivalent to that $\eta_{0}\le\eta\le1$,
which implies $\varphi$ is increasing on $[0,h_{0}]$ and decreasing
on $(h_{0},\infty)$.  Observe that $\varphi\left(0\right)=0$. Hence,
$\varphi\left(h\right)=0$ has a unique root for $h>0$.

\end{IEEEproof}
We now verify that this stationary point is the unique minimizer.
Consider a boundary point $\left(1,t\right)$. For this point, denoting
$b=D_{2}^{-1}(t)$, we have
\begin{align*}
g\left(1,t\right) & =\mathbb{D}_{2}\left(0,b\right)-\frac{1}{p}-\frac{1}{q}D_{2}\left(b\right)\\
 & =D_{2}^{(0,b)}\left(p_{0,b}^{*}\right)-\frac{1}{p}-\frac{1}{q}D_{2}\left(b\right)
\end{align*}
Denoting $a=D_{2}^{-1}(s)$, by the Taylor expansion of $g(s,t)$
at $s=1$ (for fixed $t$), we have 
\[
D_{2}^{(a,b)}\left(p_{a,b}^{*}\right)=D_{2}^{(0,b)}\left(p_{0,b}^{*}\right)+a\log a+o_{a\to0}\left(a\log a\right)
\]
and 
\[
D_{2}\left(a\right)=1+a\log a+o_{a\to0}\left(a\log a\right).
\]
Hence, 
\[
g(s,t)=g(1,t)+\left(1-\frac{1}{p}\right)a\log a+o_{a\to0}\left(a\log a\right),
\]
which implies that $\left(1,t\right)$ is not a minimizer of $g$.

Consider a boundary point $\left(0,t\right)$. For this point,
\begin{align*}
g\left(0,t\right) & =t-\frac{1}{p}t\ge0
\end{align*}
Moreover, by the information-theoretic characterization of the hypercontractivity
region \cite{nair2014equivalent}, we know that for $r<\rho^{2}$,
$\frac{x}{p}+\frac{y}{q}$ is not a supporting plane of $f$, which
means that there is a point $(x,y)$ such that $f(x,y)<\frac{x}{p}+\frac{y}{q}$.
Hence, any boundary point $\left(0,t\right)$ cannot be a minimizer. 

Combining all the cases above, we have Lemma \ref{lem:minimizers}.

\subsection{Proof of Lemma \ref{lem:convexity}}

Here we only prove the first statement. The other statement follows
similarly. 

The ``only if'' part is obvious. Here we only focus on the ``if''
part. For a function $g$, denote its epigraph as $\mathrm{epi}(g)$.
Denote $\breve{f}$ as the lower convex envelope of $f$. Let $M$
be a finite value such that $M>\max_{x\in S}f(x)$. Hence, $U:=\left(S\times(-\infty,M]\right)\cap\mathrm{epi}(\breve{f})$
is convex and compact in $\mathbb{R}^{n+1}$. By Krein--Milman theorem,
$U$ has extreme points, and if the set of extreme points is denoted
by $E\subseteq\left\{ (\mathbf{x},f(\mathbf{x})):\mathbf{x}\in S\right\} \cup\left(S\times\{M\}\right)$,
then $U$ is the closed convex hull of $E$. 

Suppose that there exists $\mathbf{x}_{0}\in S^{o}$ such that $\breve{f}(\mathbf{x}_{0})<f(\mathbf{x}_{0})$.
Then, $(\mathbf{x}_{0},\breve{f}(\mathbf{x}_{0}))$ is in the interior
of $U$, and hence is not in $E$. Hence, there exist $n+2$ points
$\mathbf{x}_{i}\in E,1\le i\le n+2$ such that $(\mathbf{x}_{0},\breve{f}(\mathbf{x}_{0}))=\sum_{i=1}^{n+2}q_{i}(\mathbf{x}_{i},\breve{f}(\mathbf{x}_{i}))$
where $\sum_{i=1}^{n+2}q_{i}=1,q_{i}\ge0,1\le i\le n+2$, and at least
two of $q_{i}\ge0,1\le i\le n+2$ are strictly positive. Let $\mathbf{x}\mapsto\left\langle \lambda,\mathbf{x}\right\rangle +c$
be a  supporting hyperplane of $\breve{f}$ at $\mathbf{x}_{0}$.
Then all the points $(\mathbf{x}_{i},\breve{f}(\mathbf{x}_{i}))$
with $q_{i}>0$ are on this hyperplane, and all of them are minimizers
of $g:\mathbf{x}\in S\mapsto f(\mathbf{x})-\left\langle \lambda,\mathbf{x}\right\rangle $.
By the assumption that the set of minimizers is convex, $\mathbf{x}_{0}$
is also a minimizer of $g$, which implies that $(\mathbf{x}_{0},f(\mathbf{x}_{0}))$
is on that hyperplane. Hence, $\breve{f}(\mathbf{x}_{0})=f(\mathbf{x}_{0})$,
contradicting with the assumption $\breve{f}(\mathbf{x}_{0})<f(\mathbf{x}_{0})$.
 Therefore, $\breve{f}(\mathbf{x}_{0})=f(\mathbf{x}_{0})$ for
all $\mathbf{x}_{0}\in S^{o}$. That is, $f$ is convex on $S^{o}$.
By the continuity of $f$, $f$ is convex on $S$.

\section{\label{sec:Proof-of-Theorem-1}Proof of Theorem \ref{thm:concavity}}

 For brevity, we denote 
\begin{align}
f\left(\alpha,\beta\right):=\widetilde{\psi}\left(\alpha,\beta\right) & =\sup_{s\leq\alpha,t\leq\beta}\mathbb{D}_{2}\left(D_{2}^{-1}\left(s\right),1-D_{2}^{-1}\left(t\right)\right)\nonumber \\
 & =\mathbb{D}_{2}\left(D_{2}^{-1}\left(\alpha\right),1-D_{2}^{-1}\left(\beta\right)\right)\label{eq:f2}
\end{align}
Denote 

\begin{align*}
\bar{f}\left(a,b\right) & =\mathbb{D}_{2}\left(a,b\right)
\end{align*}
Note that $f$ and $\bar{f}$ are differentiable. Define 
\begin{align*}
g\left(s,t\right) & :=f\left(s,t\right)-\lambda s-\mu t\\
 & =\mathbb{D}_{2}\left(D_{2}^{-1}\left(s\right),1-D_{2}^{-1}\left(t\right)\right)-\lambda s-\mu t
\end{align*}
and 
\begin{align}
\Gamma & :=\max_{s,t\in[0,1]}g\left(s,t\right)\nonumber \\
 & =\max_{a\in[0,1/2],b\in[1/2,1]}\mathbb{D}_{2}\left(a,b\right)-\lambda D_{2}\left(a\right)-\mu D_{2}\left(b\right).\label{eq:-2}
\end{align}

We have the following lemma. 
\begin{lem}
\label{lem:minimizers-2} Let $S:=[0,1]^{2}$. Let $f$ be the function
given in \eqref{eq:f2}, and $\invbreve f$ be its upper concave envelope.
Then, for any supergradient $\left(\lambda,\mu\right)$ of $\invbreve f$
at a point in $S^{o}$, the set of maximizers of the function $g:(x,y)\mapsto f(x,y)-\lambda x-\mu y$
is  a  convex subset of $S$. 
\end{lem}
Combining Lemmas \ref{lem:minimizers-2} and \ref{lem:convexity},
we have that $f$ is strictly concave. We next prove Lemma \ref{lem:minimizers-2}.

\subsection{Proof of Lemma \ref{lem:minimizers-2}}

By Lemma \ref{lem:Theta}, any supergradient $\left(\lambda,\mu\right)$
of $\invbreve f$ at a point in $S^{o}$ must satisfy that $\lambda,\mu\in[1,\infty)$.
Hence, we only need to consider $\lambda,\mu\in[1,\infty)$. In the
following, we denote $\lambda=1/p,\mu=1/q$ with $p,q\in(0,1]$. 
In the following, we denote $Q_{X}=\left\{ 1-a,a\right\} ,Q_{Y}=\left\{ 1-b,b\right\} $,
and hence, $\max_{Q_{X}}$ and $\min_{Q_{X}}$ denote optimizations
over the probability simplex $\left\{ \left(a_{0},a_{1}\right)\in\mathbb{R}_{\ge0}^{2}:a_{0}+a_{1}=1\right\} $.
Since the objective function in \eqref{eq:-2} is continuous, and
the domain of feasible solutions is compact, we know that the maximum
in \eqref{eq:-2} is attained.

In fact, \eqref{eq:-2} can be rewritten as 
\begin{align}
\Gamma & =\max_{Q_{X},Q_{Y}}\mathbb{D}\left(Q_{X},Q_{Y}\|P_{XY}\right)-\lambda D\left(Q_{X}\|P_{X}\right)-\mu D\left(Q_{Y}\|P_{Y}\right)\label{eq:-3-1}
\end{align}
where 
\begin{equation}
\mathbb{D}\left(Q_{X},Q_{Y}\|P_{XY}\right)=\min_{R_{XY}\in C(Q_{X},Q_{Y})}D(R_{XY}\|P_{XY}).\label{eq:-1}
\end{equation}
 The Lagrangian of the minimization in \eqref{eq:-1} is 
\[
L_{Q_{X},Q_{Y}}\left(\eta_{X},\eta_{Y},R_{XY}\right)=D\left(R_{XY}\|P_{XY}\right)+\sum_{x}\eta_{X}(x)\left(R_{X}(x)-Q_{X}(x)\right)+\sum_{y}\eta_{Y}(y)\left(R_{Y}(y)-Q_{Y}(y)\right).
\]
Since the minimization in \eqref{eq:-1} is a convex optimization
problem with linear constraints, Slater\textquoteright s condition
is satisfied, which in turn implies that the strong duality holds
and the optimal solution of the dual problem exists \cite{boyd2004convex}.
Hence, 
\begin{align*}
\mathbb{D}\left(Q_{X},Q_{Y}\|P_{XY}\right) & =\max_{\eta_{X},\eta_{Y}}\min_{R_{XY}}L_{Q_{X},Q_{Y}}\left(\eta_{X},\eta_{Y},R_{XY}\right)
\end{align*}
and the maximum is attained. Substituting this into \eqref{eq:-3-1}
yields 

\begin{align}
\Gamma & =\max_{Q_{X},Q_{Y},\eta_{X},\eta_{Y}}\min_{R_{XY}}K\left(Q_{X},Q_{Y},\eta_{X},\eta_{Y},R_{XY}\right)\label{eq:-14}
\end{align}
where 
\[
K\left(Q_{X},Q_{Y},\eta_{X},\eta_{Y},R_{XY}\right):=L_{Q_{X},Q_{Y}}\left(\eta_{X},\eta_{Y},R_{XY}\right)-\lambda D\left(Q_{X}\|P_{X}\right)-\mu D\left(Q_{Y}\|P_{Y}\right).
\]
Let $\left(Q_{X}^{*},Q_{Y}^{*},\eta_{X}^{*},\eta_{Y}^{*}\right)$
be a maximizer in \eqref{eq:-14}, and given $\left(Q_{X}^{*},Q_{Y}^{*},\eta_{X}^{*},\eta_{Y}^{*}\right)$,
$R_{XY}^{*}$ is a minimizer for the inner minimization. Observe that
\begin{align*}
\Gamma & =\max_{Q_{X},Q_{Y}}\min_{R_{XY}}K\left(Q_{X},Q_{Y},\eta_{X}^{*},\eta_{Y}^{*},R_{XY}\right)\\
 & =\min_{R_{XY}}\left\{ D\left(R_{XY}\|P_{XY}\right)+\sum_{x}\eta_{X}^{*}(x)R_{X}(x)+\sum_{y}\eta_{Y}^{*}(y)R_{Y}(y\right\} \\
 & \qquad-\min_{Q_{X},Q_{Y}}\left\{ \lambda D\left(Q_{X}\|P_{X}\right)+\mu D\left(Q_{Y}\|P_{Y}\right)+\sum_{x}\eta_{X}^{*}(x)Q_{X}(x)+\sum_{y}\eta_{Y}^{*}(y)Q_{Y}(y)\right\} .
\end{align*}
By Lagrangian conditions, 
\begin{align}
R_{XY}^{*}\left(x,y\right) & =\frac{P_{XY}\left(x,y\right)e^{-\eta_{X}^{*}(x)-\eta_{Y}^{*}(y)}}{\sum_{x,y}P_{XY}\left(x,y\right)e^{-\eta_{X}^{*}(x)-\eta_{Y}^{*}(y)}}\label{eq:-19}\\
Q_{X}^{*}(x) & =\frac{P_{X}\left(x\right)e^{-\eta_{X}^{*}(x)/\lambda}}{\sum_{x}P_{X}\left(x\right)e^{-\eta_{X}^{*}(x)/\lambda}}\\
Q_{Y}^{*}(y) & =\frac{P_{Y}\left(y\right)e^{-\eta_{Y}^{*}(y)/\mu}}{\sum_{y}P_{Y}\left(y\right)e^{-\eta_{Y}^{*}(y)/\mu}}.\label{eq:-20}
\end{align}

Observe that $K\left(Q_{X},Q_{Y},\eta_{X},\eta_{Y},R_{XY}\right)$
is convex in $R_{XY}$, and concave in $\left(\eta_{X},\eta_{Y}\right)$,
by the strong duality, 
\begin{equation}
\Gamma=\min_{R_{XY}}\max_{\eta_{X},\eta_{Y}}K\left(Q_{X}^{*},Q_{Y}^{*},\eta_{X},\eta_{Y},R_{XY}\right).\label{eq:-15}
\end{equation}
Let $R_{XY}^{**}$ be a minimizer for the minimization in \eqref{eq:-15}.
Given $\left(Q_{X}^{*},Q_{Y}^{*}\right)$, by the strong duality,
$\left(\eta_{X}^{*},\eta_{Y}^{*},R_{XY}^{**}\right)$ is a saddle
point of $\left(\eta_{X},\eta_{Y},R_{XY}\right)\mapsto K\left(Q_{X}^{*},Q_{Y}^{*},\eta_{X},\eta_{Y},R_{XY}\right)$,
which hence satisfies 
\begin{align}
R_{XY}^{**}\left(x,y\right) & =\frac{P_{XY}\left(x,y\right)e^{-\eta_{X}^{*}(x)-\eta_{Y}^{*}(y)}}{\sum_{x,y}P_{XY}\left(x,y\right)e^{-\eta_{X}^{*}(x)-\eta_{Y}^{*}(y)}}\label{eq:-17}\\
R_{X}^{**} & =Q_{X}^{*}\\
R_{Y}^{**} & =Q_{Y}^{*}.\label{eq:-21}
\end{align}
Comparing \eqref{eq:-19} with \eqref{eq:-17} we have $R_{XY}^{*}=R_{XY}^{**}$.
Therefore, from \eqref{eq:-19}-\eqref{eq:-20} and \eqref{eq:-17}-\eqref{eq:-21},
we have
\begin{align*}
R_{XY}^{*}\left(x,y\right) & =\frac{P_{XY}\left(x,y\right)e^{-\eta_{X}^{*}(x)-\eta_{Y}^{*}(y)}}{\sum_{x,y}P_{XY}\left(x,y\right)e^{-\eta_{X}^{*}(x)-\eta_{Y}^{*}(y)}}\\
R_{X}^{*} & =Q_{X}^{*}\\
R_{Y}^{*} & =Q_{Y}^{*}\\
Q_{X}^{*}(x) & =\frac{P_{X}\left(x\right)e^{-\eta_{X}^{*}(x)/\lambda}}{\sum_{x}P_{X}\left(x\right)e^{-\eta_{X}^{*}(x)/\lambda}}\\
Q_{Y}^{*}(y) & =\frac{P_{Y}\left(y\right)e^{-\eta_{Y}^{*}(y)/\mu}}{\sum_{y}P_{Y}\left(y\right)e^{-\eta_{Y}^{*}(y)/\mu}}.
\end{align*}

To prove the uniqueness of optimal $\left(Q_{X}^{*},Q_{Y}^{*}\right)$,
it suffices to show the uniqueness of the solution of the following
equations with $R_{XY}$ unknown. 
\begin{equation}
R_{XY}\left(x,y\right)=\frac{\hat{\Pi}\left(x,y\right)}{\sum_{x,y\in\{0,1\}}\hat{\Pi}\left(x,y\right)},\forall x,y\in\{0,1\}\label{eq:-7}
\end{equation}
where
\[
\hat{\Pi}\left(x,y\right)=P_{XY}\left(x,y\right)\left(\frac{R_{X}(x)}{P_{X}(x)}\right)^{\lambda}\left(\frac{R_{Y}(y)}{P_{Y}(y)}\right)^{\mu}.
\]

Denote $R_{XY}:=\begin{bmatrix}q_{00} & q_{01}\\
q_{10} & q_{11}
\end{bmatrix}$. From \eqref{eq:-7}, we have \eqref{eq:-90} and \eqref{eq:-91}
but with $\frac{q_{00}+q_{01}}{q_{10}+q_{11}}\ge1$ and $\frac{q_{00}+q_{10}}{q_{01}+q_{11}}\le1$.

If $p=1,q\in(0,1]$, then solving \eqref{eq:-90} and \eqref{eq:-91},
we have $\left(q_{00},q_{01},q_{10},q_{11}\right)=\left(\frac{1+\rho}{4},\frac{1-\rho}{4},\frac{1-\rho}{4},\frac{1+\rho}{4}\right)$.
Obviously, this stationary point is not a maximizer. Hence, all the
maximizers are on the boundary $s=1$. For this case, 
\begin{align}
g\left(1,t\right) & =f\left(1,t\right)-1-\frac{t}{q}\nonumber \\
 & =\left(1-\frac{1}{q}\right)D_{2}(b)-b\log\frac{1-\rho}{2}-(1-b)\log\frac{1+\rho}{2},\label{eq:-8}
\end{align}
where $b=D_{2}^{-1}(t)$. If $q\in(0,1)$, then \eqref{eq:-8} is
strictly concave in $b$. Hence, the maximizer of $g$ is unique for
this case. If $q=1$, then \eqref{eq:-8} is maximized uniquely at
$b=1$, and hence, the maximizer of $g$ is also unique (i.e., $s=t=1$)
for this case. Hence, the maximizer of $g$ is unique for $p=1,q\in(0,1]$.
By symmetry, the maximizer of $g$ is also unique for $q=1,p\in(0,1]$.

It remains to consider the case $p,q\in(0,1)$. Denote $r=\left(p-1\right)\left(q-1\right)$.
If $r>\rho^{2}$ and $p,q\in(0,1)$, then by the information-theoretic
characterization of the reverse hypercontractivity region, $(0,0)$
is the unique maximizer of $g$. Hence, Lemma \ref{lem:minimizers}
is satisfied for this case.

We next consider the case $0<r\le\rho^{2},p,q\in(0,1)$. Denote $\theta=\frac{1-\rho}{1+\rho}$,
$u:=\frac{1}{p-1},v:=\frac{1}{q-1}$. For this case, we have $u,v<-1$.
 Solving \eqref{eq:-90} and \eqref{eq:-91} , we have \eqref{eq:-95}.
By Lemma \ref{lem:uniqueness}, there is only one stationary point
of $g$. 

We now verify that this stationary point is the unique maximizer.
By the information-theoretic characterization of the reverse hypercontractivity
region, $\Gamma=\max_{s,t\in[0,1]}g\left(s,t\right)$ is positive.
Hence, $(0,0)$ is not a maximizer of $g$. We next consider a boundary
point $\left(1,t\right)$. Denoting $a=D_{2}^{-1}(s),b=D_{2}^{-1}(t)$,
by the Taylor expansion of $f(s,t)$ at $s=1$ (for fixed $t$), we
have 
\[
D_{2}^{(a,b)}\left(p_{a,b}^{*}\right)=D_{2}^{(0,b)}\left(p_{0,b}^{*}\right)+a\log a+o_{a\to0}\left(a\log a\right)
\]
and 
\[
D_{2}\left(a\right)=1+a\log a+o_{a\to0}\left(a\log a\right).
\]
That is, $\frac{\partial}{\partial s}f(s,t)|_{s=1}=1$ for any $t\in[0,1]$.
Hence, 
\[
g(s,t)=g(1,t)+\left(1-\frac{1}{p}\right)a\log a+o_{a\to0}\left(a\log a\right),
\]
which implies that $\left(1,t\right)$ is not a maximizer of $g$.
 Similarly, any point $(s,1)$ is also not a maximizer of $g$. 

Consider a boundary point $\left(0,t\right)$. For this point, $\frac{\partial}{\partial s}f(s,t)|_{s=0}=\infty$
for all $t\in[0,1]$. Hence, any point $\left(0,t\right)$ is not
a maximizer of $g$. Similarly, any point $(s,0)$ is also not a maximizer
of $g$. 

Combining all the cases above, we have Lemma \ref{lem:minimizers}.

\section{\label{sec:Proof-of-Theorem-1-1}Proof of Theorem \ref{thm:concavity2}}

For brevity, we denote 
\begin{align}
f\left(s\right):=\varphi_{q}\left(s\right) & =\min_{0\le t\le1}\varphi\left(s,t\right)-\frac{t}{q}\nonumber \\
 & =\min_{0\le t\le1}\mathbb{D}_{2}\left(D_{2}^{-1}\left(s\right),D_{2}^{-1}\left(t\right)\right)-\frac{t}{q}.\label{eq:f2-1}
\end{align}
Denote 

\begin{align*}
\bar{f}\left(a,b\right) & =\mathbb{D}_{2}\left(a,b\right)
\end{align*}
Note that $f$ and $\bar{f}$ are differentiable. Define 
\begin{align*}
g\left(s\right) & :=f\left(s\right)-\lambda s
\end{align*}
and 
\begin{align}
\Gamma & :=\max_{s\in[0,1]}g\left(s\right)\nonumber \\
 & =\max_{a\in[0,1/2]}\min_{b\in[0,1/2]}\mathbb{D}_{2}\left(a,b\right)-\frac{D_{2}\left(b\right)}{q}-\lambda D_{2}\left(a\right).\label{eq:-2-1}
\end{align}

We have the following lemma. 
\begin{lem}
\label{lem:minimizers-2-1} Let $S:=[0,1]^{2}$. Let $f$ be the
function given in \eqref{eq:f2-1}, and $\invbreve f$ be its upper
concave envelope. Then, for any supergradient $\left(\lambda,\mu\right)$
of $\invbreve f$ at a point in $S^{o}$, the set of maximizers of
the function $g:(x,y)\mapsto f(x,y)-\lambda x-\mu y$ is  a  convex
subset of $S$. 
\end{lem}
Combining Lemmas \ref{lem:minimizers-2-1} and \ref{lem:convexity},
we have that $f$ is strictly concave. We next prove Lemma \ref{lem:minimizers-2-1}. 

\subsection{Proof of Lemma \ref{lem:minimizers-2-1}}

By Lemma \ref{lem:Theta-1}, any supergradient $\lambda$ of $\invbreve f$
at a point in $S^{o}$ must satisfy that $\lambda\in[1,\infty)$.
In the following, we denote $\lambda=1/p,\mu=1/q$ with $p\in(0,1],q<0$.

In fact, \eqref{eq:-2-1} can be rewritten as 
\begin{align}
\Gamma & =\max_{Q_{X}}\min_{Q_{Y}}\mathbb{D}\left(Q_{X},Q_{Y}\|P_{XY}\right)-\mu D\left(Q_{Y}\|P_{Y}\right)-\lambda D\left(Q_{X}\|P_{X}\right)\label{eq:-3-1-1}\\
 & =\max_{Q_{X}}\left\{ \min_{Q_{Y},R_{XY}:R_{X}=Q_{X},R_{Y}=Q_{Y}}\left\{ D\left(R_{XY}\|P_{XY}\right)-\mu D\left(Q_{Y}\|P_{Y}\right)\right\} -\lambda D\left(Q_{X}\|P_{X}\right)\right\} \label{eq:-11}
\end{align}
where $\mathbb{D}\left(Q_{X},Q_{Y}\|P_{XY}\right)$ is defined in
\eqref{eq:-1}. The Lagrangian of the inner minimization in \eqref{eq:-11}
is 
\begin{align*}
L_{Q_{X}}\left(Q_{Y},R_{XY},\eta_{X},\eta_{Y}\right) & =D\left(R_{XY}\|P_{XY}\right)-\mu D\left(Q_{Y}\|P_{Y}\right)+\sum_{x}\eta_{X}(x)\left(R_{X}(x)-Q_{X}(x)\right)\\
 & \qquad+\sum_{y}\eta_{Y}(y)\left(R_{Y}(y)-Q_{Y}(y)\right).
\end{align*}
 Since the inner minimization in \eqref{eq:-11} is a convex optimization
problem with linear constraints, the strong duality holds. Hence,
the inner minimization in \eqref{eq:-11} is equal to 
\begin{align*}
 & \max_{\eta_{X},\eta_{Y}}\min_{Q_{Y},R_{XY}}L_{Q_{X}}\left(Q_{Y},R_{XY},\eta_{X},\eta_{Y}\right).
\end{align*}
Substituting this into \eqref{eq:-11} yields 
\begin{align}
\Gamma & =\max_{Q_{X},\eta_{X},\eta_{Y}}\min_{Q_{Y},R_{XY}}K\left(Q_{X},Q_{Y},\eta_{X},\eta_{Y},R_{XY}\right)\label{eq:-14-1}
\end{align}
where 
\[
K\left(Q_{X},Q_{Y},\eta_{X},\eta_{Y},R_{XY}\right):=L_{Q_{X}}\left(Q_{Y},R_{XY},\eta_{X},\eta_{Y}\right)-\lambda D\left(Q_{X}\|P_{X}\right).
\]
Let $\left(Q_{X}^{*},\eta_{X}^{*},\eta_{Y}^{*}\right)$ be a maximizer
in \eqref{eq:-14-1}, and given $\left(Q_{X}^{*},\eta_{X}^{*},\eta_{Y}^{*}\right)$,
$\left(Q_{Y}^{*},R_{XY}^{*}\right)$ is a minimizer for the inner
minimization. Observe that 
\begin{align*}
\Gamma & =\max_{Q_{X}}\min_{Q_{Y},R_{XY}}K\left(Q_{X},Q_{Y},\eta_{X}^{*},\eta_{Y}^{*},R_{XY}\right)\\
 & =\min_{R_{XY}}\left\{ D\left(R_{XY}\|P_{XY}\right)+\sum_{x}\eta_{X}^{*}(x)R_{X}(x)+\sum_{y}\eta_{Y}^{*}(y)R_{Y}(y\right\} \\
 & \qquad-\min_{Q_{X}}\left\{ \lambda D\left(Q_{X}\|P_{X}\right)+\sum_{x}\eta_{X}^{*}(x)Q_{X}(x)\right\} \\
 & \qquad-\max_{Q_{Y}}\left\{ \mu D\left(Q_{Y}\|P_{Y}\right)+\sum_{y}\eta_{Y}^{*}(y)Q_{Y}(y)\right\} .
\end{align*}
By Lagrangian conditions, 
\begin{align}
R_{XY}^{*}\left(x,y\right) & =\frac{P_{XY}\left(x,y\right)e^{-\eta_{X}^{*}(x)-\eta_{Y}^{*}(y)}}{\sum_{x,y}P_{XY}\left(x,y\right)e^{-\eta_{X}^{*}(x)-\eta_{Y}^{*}(y)}}\label{eq:-19-1}\\
Q_{X}^{*}(x) & =\frac{P_{X}\left(x\right)e^{-\eta_{X}^{*}(x)/\lambda}}{\sum_{x}P_{X}\left(x\right)e^{-\eta_{X}^{*}(x)/\lambda}}\\
Q_{Y}^{*}(y) & =\frac{P_{Y}\left(y\right)e^{-\eta_{Y}^{*}(y)/\mu}}{\sum_{y}P_{Y}\left(y\right)e^{-\eta_{Y}^{*}(y)/\mu}}.\label{eq:-20-1}
\end{align}

Observe that $K\left(Q_{X},Q_{Y},\eta_{X},\eta_{Y},R_{XY}\right)$
is convex in $\left(Q_{Y},R_{XY}\right)$, and concave in $\left(\eta_{X},\eta_{Y}\right)$,
by the strong duality, 
\begin{equation}
\Gamma=\min_{Q_{Y},R_{XY}}\max_{\eta_{X},\eta_{Y}}K\left(Q_{X}^{*},Q_{Y},\eta_{X},\eta_{Y},R_{XY}\right).\label{eq:-15-1}
\end{equation}
Let $\left(Q_{Y}^{**},R_{XY}^{**}\right)$ be a minimizer for the
minimization in \eqref{eq:-15-1}. Given $Q_{X}^{*}$, by the strong
duality, $\left(\eta_{X}^{*},\eta_{Y}^{*},R_{XY}^{**}\right)$ is
a saddle point of $\left(\eta_{X},\eta_{Y},Q_{Y},R_{XY}\right)\mapsto K\left(Q_{X}^{*},Q_{Y},\eta_{X},\eta_{Y},R_{XY}\right)$,
which hence satisfies 
\begin{align}
R_{XY}^{**}\left(x,y\right) & =\frac{P_{XY}\left(x,y\right)e^{-\eta_{X}^{*}(x)-\eta_{Y}^{*}(y)}}{\sum_{x,y}P_{XY}\left(x,y\right)e^{-\eta_{X}^{*}(x)-\eta_{Y}^{*}(y)}}\label{eq:-17-1}\\
R_{X}^{**} & =Q_{X}^{*}\\
R_{Y}^{**} & =Q_{Y}^{**}\label{eq:-21-1}\\
Q_{Y}^{**}(y) & =\frac{P_{Y}\left(y\right)e^{-\eta_{Y}^{*}(y)/\mu}}{\sum_{y}P_{Y}\left(y\right)e^{-\eta_{Y}^{*}(y)/\mu}}.
\end{align}
Comparing \eqref{eq:-19-1} with \eqref{eq:-17-1} we have $R_{XY}^{*}=R_{XY}^{**},Q_{Y}^{*}=Q_{Y}^{**}$.
Therefore, from \eqref{eq:-19-1}-\eqref{eq:-20-1} and \eqref{eq:-17-1}-\eqref{eq:-21-1},
we have
\begin{align*}
R_{XY}^{*}\left(x,y\right) & =\frac{P_{XY}\left(x,y\right)e^{-\eta_{X}^{*}(x)-\eta_{Y}^{*}(y)}}{\sum_{x,y}P_{XY}\left(x,y\right)e^{-\eta_{X}^{*}(x)-\eta_{Y}^{*}(y)}}\\
R_{X}^{*} & =Q_{X}^{*}\\
R_{Y}^{*} & =Q_{Y}^{*}\\
Q_{X}^{*}(x) & =\frac{P_{X}\left(x\right)e^{-\eta_{X}^{*}(x)/\lambda}}{\sum_{x}P_{X}\left(x\right)e^{-\eta_{X}^{*}(x)/\lambda}}\\
Q_{Y}^{*}(y) & =\frac{P_{Y}\left(y\right)e^{-\eta_{Y}^{*}(y)/\mu}}{\sum_{y}P_{Y}\left(y\right)e^{-\eta_{Y}^{*}(y)/\mu}}.
\end{align*}
 To prove the uniqueness of $Q_{X}^{*}$, it suffices to show
the uniqueness of the solution of the following equations with $R_{XY}$
unknown. 
\begin{equation}
R_{XY}\left(x,y\right)=\frac{\hat{\Pi}\left(x,y\right)}{\sum_{x,y\in\{0,1\}}\hat{\Pi}\left(x,y\right)},\forall x,y\in\{0,1\}\label{eq:-7-1}
\end{equation}
where
\[
\hat{\Pi}\left(x,y\right)=P_{XY}\left(x,y\right)\left(\frac{R_{X}(x)}{P_{X}(x)}\right)^{\lambda}\left(\frac{R_{Y}(y)}{P_{Y}(y)}\right)^{\mu}.
\]

Denote $R_{XY}:=\begin{bmatrix}q_{00} & q_{01}\\
q_{10} & q_{11}
\end{bmatrix}$. From \eqref{eq:-7-1}, we have \eqref{eq:-90} and \eqref{eq:-91}
with $\frac{q_{00}+q_{01}}{q_{10}+q_{11}}\ge1$ and $\frac{q_{00}+q_{10}}{q_{01}+q_{11}}\ge1$.
However, here we denote $z$ as the expressions in \eqref{eq:-91}
(rather than the ones in \eqref{eq:-90}).

If $p=1,q<0$, then solving \eqref{eq:-90} and \eqref{eq:-91}, we
have $\left(q_{00},q_{01},q_{10},q_{11}\right)=\left(\frac{1+\rho}{4},\frac{1-\rho}{4},\frac{1-\rho}{4},\frac{1+\rho}{4}\right)$.
This is the boundary point $s=0$ and $g\left(0\right)=0$. For
the boundary $s=1$, 
\begin{align}
g\left(1\right) & =\min_{0\le t\le1}\varphi\left(1,t\right)-\frac{t}{q}-1\nonumber \\
 & =\min_{0\le b\le1/2}\mathbb{D}_{2}\left(0,b\right)-\frac{D_{2}(b)}{q}-1\\
 & =\min_{0\le b\le1/2}\left(1-\frac{1}{q}\right)D_{2}(b)-b\log\frac{1-\rho}{2}-(1-b)\log\frac{1+\rho}{2},\label{eq:-8-1}
\end{align}
where $b=D_{2}^{-1}(t)$. Obviously, $g\left(1\right)>0$. For $q<0$,
\eqref{eq:-8-1} is strictly convex in $b$. Hence, the maximizer
of $g$ is unique. 

It remains to consider the case $p\in(0,1),q<0$. Denote $r=\left(p-1\right)\left(q-1\right)$.
If $r>\rho^{2}$ and $p,q\in(0,1)$, then by the information-theoretic
characterization of the reverse hypercontractivity region \cite{beigi2016equivalent},
$(0,0)$ is the unique maximizer of $g$. Hence, Lemma \ref{lem:minimizers}
is satisfied for this case.

We next consider the case $0<r\le\rho^{2},p\in(0,1),q<0$. Denote
$\theta=\frac{1-\rho}{1+\rho}$, $u:=\frac{1}{p-1},v:=\frac{1}{q-1}$.
For this case, we have $u<-1$.   Solving \eqref{eq:-90} and \eqref{eq:-91}
, we have \eqref{eq:-95}. By Lemma \ref{lem:uniqueness}, there
is only one stationary point of $g$. 

We now verify that this stationary point, denoted as $s^{*}$, or
the boundary point $s=1$ is the unique maximizer. By the information-theoretic
characterization of the reverse hypercontractivity region \cite{beigi2016equivalent},
$\Gamma=\max_{s\in[0,1]}g\left(s\right)$ is positive. This implies
that the boundary point $s=0$ is not optimal since $g\left(0\right)=0$.
Moreover, if both the stationary point $s^{*}$ and the boundary point
$s=1$ are maximizers, then there must exist a local minimum $\hat{s}^{*}$
of $g$ which is strictly between $s^{*}$ and $1$. From the definition
of $g$, we know that there exists some $\hat{t}^{*}$ such that $\left(\hat{s}^{*},\hat{t}^{*}\right)$
is a local minimum of $\hat{g}:\left(s,t\right)\mapsto\varphi\left(s,t\right)-\frac{t}{q}-\frac{s}{p}$.
Moreover, given $q<0$ and $0<a<1/2$, $b\mapsto\mathbb{D}_{2}\left(a,b\right)-\frac{D_{2}(b)}{q}$
is convex, and the minimum of $b\mapsto\mathbb{D}_{2}\left(a,b\right)-\frac{D_{2}(b)}{q}$
is attained at a point strictly between $0<b<1/2$. Hence, $0<\hat{t}^{*}<1$,
which implies that $\left(\hat{s}^{*},\hat{t}^{*}\right)$ is a stationary
point of $\hat{g}$.  From the proof of Lemma \ref{lem:minimizers},
the distribution $\left(q_{00},q_{01},q_{10},q_{11}\right)$ induced
$\left(\hat{s}^{*},\hat{t}^{*}\right)$ must satisfy Lagrangian conditions
in \eqref{eq:-90} and \eqref{eq:-91}. This further implies that
the $z$ defined as the expressions in \eqref{eq:-91} induced by
$\left(\hat{s}^{*},\hat{t}^{*}\right)$ satisfy \eqref{eq:-95}, and
it is different from the $z$ induced by $s^{*}$ since $\hat{s}^{*}\neq s^{*}$.
This contradicts with Lemma \ref{lem:uniqueness}. Therefore, the
stationary point $s^{*}$ or the boundary point $s=1$ is the unique
maximizer of $g$. This completes the proof of Lemma \ref{lem:minimizers}.

\bibliographystyle{unsrt}
\bibliography{ref}

\begin{thebibliography}{10}

\bibitem{yu2021Graphs}
L.~Yu, V.~Anantharam, and J.~Chen.
\newblock Graphs of joint types, noninteractive simulation,and stronger
  hypercontractivity.
\newblock {\em arXiv preprint arXiv:2102.00668}, Feb. 2021. [Online].
  Available: https://arxiv.org/abs/2102.00668.

\bibitem{yu2021strong}
L.~Yu.
\newblock Strong brascamp-lieb inequalities.
\newblock {\em arXiv preprint arXiv:2102.06935}, 2021.

\bibitem{ordentlich2020note}
O.~Ordentlich, Y.~Polyanskiy, and O.~Shayevitz.
\newblock A note on the probability of rectangles for correlated binary
  strings.
\newblock {\em IEEE Trans. Inf. Theory}, 2020.

\bibitem{kirshner2019moment}
N.~Kirshner and A.~Samorodnitsky.
\newblock A moment ratio bound for polynomials and some extremal properties of
  {Krawchouk} polynomials and {Hamming} spheres.
\newblock {\em arXiv preprint arXiv:1909.11929}, 2019.

\bibitem{polyanskiy2019hypercontractivity2}
Y.~Polyanskiy.
\newblock Hypercontractivity for sparse functions on the discrete hypercube.
\newblock {\em manuscript}, 2019.

\bibitem{Gamal}
A.~El Gamal and Y.-H. Kim.
\newblock {\em Network Information Theory}.
\newblock Cambridge University Press, 2011.

\bibitem{nair2014equivalent}
C.~Nair.
\newblock Equivalent formulations of hypercontractivity using information
  measures.
\newblock In {\em International Zurich Seminar}, 2014.

\bibitem{nair2016evaluating}
C.~Nair and Y.~N. Wang.
\newblock Evaluating hypercontractivity parameters using information measures.
\newblock In {\em 2016 IEEE International Symposium on Information Theory
  (ISIT)}, pages 570--574. IEEE, 2016.

\bibitem{boyd2004convex}
S.~Boyd and L.~Vandenberghe.
\newblock {\em Convex optimization}.
\newblock Cambridge university press, 2004.

\bibitem{beigi2016equivalent}
S.~Beigi and C.~Nair.
\newblock Equivalent characterization of reverse {Brascamp-Lieb-type}
  inequalities using information measures.
\newblock In {\em 2016 IEEE International Symposium on Information Theory
  (ISIT)}, pages 1038--1042. IEEE, 2016.

\end{thebibliography}

\end{document}